\documentclass[a4paper]{article}
\usepackage{hiph-art}
\usepackage{graphicx}
\usepackage{amssymb}
\volnumber{19} \issuenumber{1} \edyear{2004}                             %|
\frompage{000} \topage{000}                                              %|
\recrevdate{31 March 2005}                                               %|
\def\rightmark{Hadronic physics with the CMS experiment}
\def\leftmark{F. Sikl\'er for the CMS Collaboration}
 
\title{Hadronic physics with the CMS experiment} 
\authors{ 
{Ferenc Sikl\'er$^1$ for the CMS Collaboration %
\index{Sikl\'er, F.}
}\\[2.812mm]
{\normalsize
\hspace*{-8pt}$^1$ KFKI Research Institute for Particle and Nuclear Physics,\\ 
1121 Budapest, Hungary\\[0.2ex]
}}
 
\abstract{The capabilities of the CMS detector are shown and its
Heavy Ion program is outlined.}



\def\pt{$\mathrm{p_{_T}}$}
\def\Et{$\mathrm{E_{_T}}$}
 
\makeindex
\begin{document}
 
\maketitle

\section{Introduction}

Experiments observing nucleus-nucleus collisions have been studying hot
nuclear matter for several decades at ultra-relativistic energies. With a
long history (BNL AGS, CERN SPS, BNL RHIC) the next step is again at CERN
with the Large Hadron Collider (LHC).

The field has recently brought exciting results, such as anomalous
suppression of heavy mesons at the SPS \cite{jpsi1,jpsi2}, the suppression
of high \pt~jets \cite{ptsupp1,ptsupp2,ptsupp3} and the disappearance of
back-to-back high \pt~hadron correlations at RHIC \cite{jetcorr},
ultimately leading to the observation of a new kind of strongly
interacting matter.

In the coming LHC era the higher available energy (5.5 TeV per nucleon
pair for A+A) will provide a wider kinematic range and higher
cross-sections for rare probes. Familiar as well as new observables will
be available: enhancement of particles with high \pt; the
"temperature"-indicators $J/\psi$, $\psi'$ and $\Upsilon$ family; medium
effects on jets, such as their shape and fragmentation; dijets, the newly
available jet-$\gamma$, jet-$\mathrm{Z^0}$ correlations; the centrality
dependence of these effects.

\section{Heavy ion experiments at LHC}

The LHC is being built in the LEP tunnel. It will provide collisions for
four experiments, three of them which have a specific heavy ion program.
One of them is the Compact Muon Solenoid (CMS) detector which has good
acceptance, spatial and momentum resolution, and is able to clearly
observe rare signals. CMS is somewhat complementary to the other heavy ion
experiments. While ALICE has particle identification capability at
moderate \pt, CMS is good at observing higher \pt~particles, especially
muons, and jets.

The CMS Heavy Ion component is an integral part of the physics program of
the experiment with both a detailed A+A and p+A program \cite{baur}.

\begin{figure}
  \includegraphics[width=\linewidth]{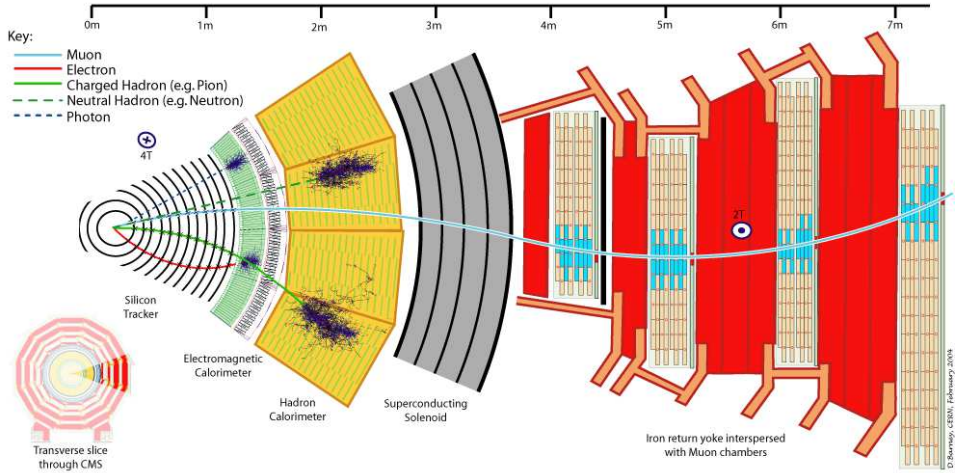}
  \caption{\label{fig:path} Slice through CMS showing particles incident 
on the different sub-detectors.}
\end{figure}

\section{The CMS detector}

The detection of charged and neutral particles, leptons and hadrons is
achieved using several detector components: namely the silicon tracker,
electromagnetic and hadronic calorimeters, muon chambers
(Fig.~\ref{fig:path}).

The silicon tracker includes pixels and strips, it has geometric coverage
for $\vert\eta\vert < 2.5$. There are about 10 million microstrips and 40
million pixels with a size of $100 \times 150 \mu\mathrm{m}^2$. The pixel
part consists of three barrel and two forward layers on each side. The
occupancy of the pixels is expected to be at the few percent level for a
multiplicity of $dN/d\eta = 5000$, even for the innermost layer at a
radius of 4.5 cm. The strip part consists of ten barrel and nine forward
layers on each side, some of which are single-, others are double-sided.
The silicon tracker has excellent reconstruction performance for \pt $>$ 1
GeV/$c$. For lower \pt~the reconstruction capabilities are limited by the
high magnetic field and effects of the material in the detector.

The electromagnetic calorimeter is very compact: its lead tungstate
crystals have high density, small Moli\'ere radius and short radiation
length. It includes barrel and endcap parts covering $\vert\eta\vert < 3$
with a granularity of $2\times 2 \mathrm{cm}^2$, the resolution is
$2.7\%/\sqrt{E} \oplus 0.55\%$. The hadronic calorimeter has barrel,
endcap and forward parts covering $\vert\eta\vert < 5$, with a resolution
of $116\%/\sqrt{E} \oplus 5\%$. The barrel and endcap are made of copper
absorber plates with scintillator sheets. The forward part is made of iron
with quartz fibers for Cerenkov-light detection. A forward calorimeter
(CASTOR) can extend the pseudorapidity range up to 7.

The muon system provides an important input to the heavy ion physics
program. Drift tubes are used in the barrel part for $\vert\eta\vert <
1.5$, cathode strip chambers are employed in endcaps up to pseudorapidity
of 2.5. Thus, the coverage is outside the fragmentation regions and
complements ALICE, which detects muons in the range $2.5 < \eta < 4.5$.

In the very forward direction a zero degree calorimeter will be used to
detect spectator neutrons (and forward photons). This information
essential for the determination of the centrality of the A+A collisions.

The high luminosity and collision rate at the LHC requires a good event
selection and powerful data acquisition system. (The expected size of
Pb+Pb events in CMS is 2-3 Mbytes.) The system consists of a low level
(hardware)  and a high level (software) trigger: the latter can already
make use of tracking information. The high level trigger system has to use
around 1000 CPUs to make fast enough decisions, which enable the storage
of interesting events at the rate of 40 Hz.

\begin{figure}
 \begin{minipage}[b]{0.55\linewidth}
  \centering
  \includegraphics[width=\linewidth]{tracks_new}

  \caption{\label{fig:tracking} Efficiency (higher lines) and fake track
rate (lower lines) for A+A collisions at $dN/dy = 1000$ (solid) and $dN/dy
= 5000$ (dashed) as a function of \pt.}

 \end{minipage}
 \begin{minipage}[b]{0.45\linewidth}
  \centering
  \includegraphics[width=\linewidth]{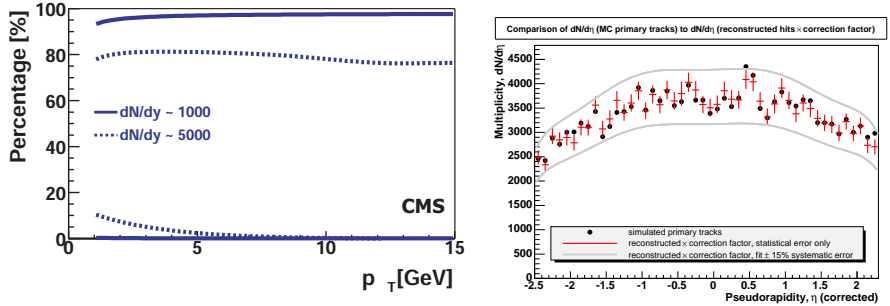}

  \caption{\label{fig:multi} Comparison of the reconstructed $dN/d\eta$
distribution to the simulated primary tracks as a function of $\eta$.}

 \end{minipage}
\end{figure}

For offline computation enormous resources are needed: a world-wide
computing grid of many thousands of CPUs is built up (LHC Computing Grid).

\begin{figure}
 \begin{minipage}[b]{0.5\linewidth}
  \centering
  \includegraphics[width=\linewidth]{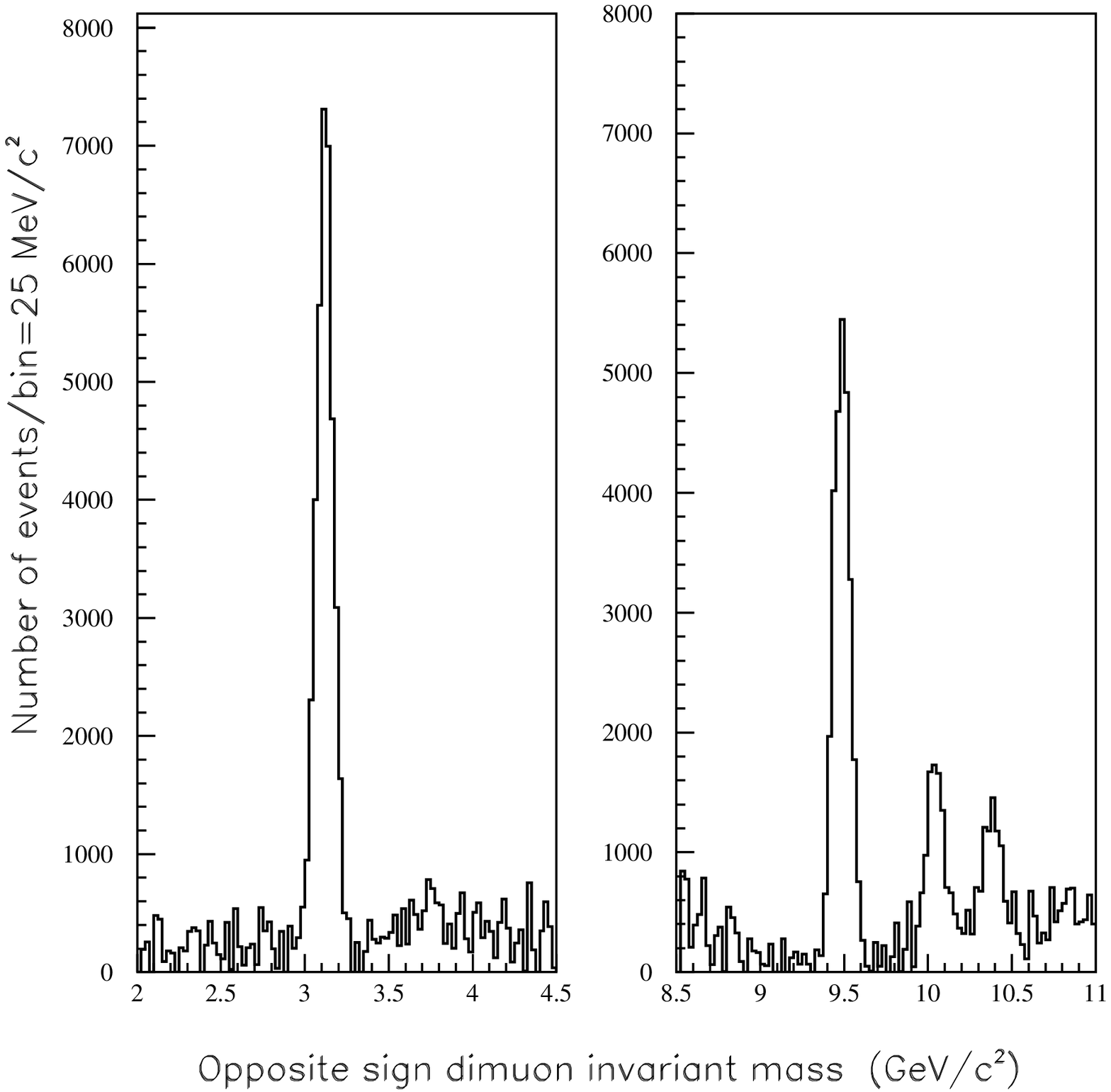}

  \caption{\label{fig:quarkonia} Mass spectrum of opposite sign dimouns
after background subtraction, with clear signs of $J/\psi$ and the
$\Upsilon$ family.}

 \end{minipage}
 \begin{minipage}[b]{0.5\linewidth}
  \centering
  \includegraphics[width=\linewidth]{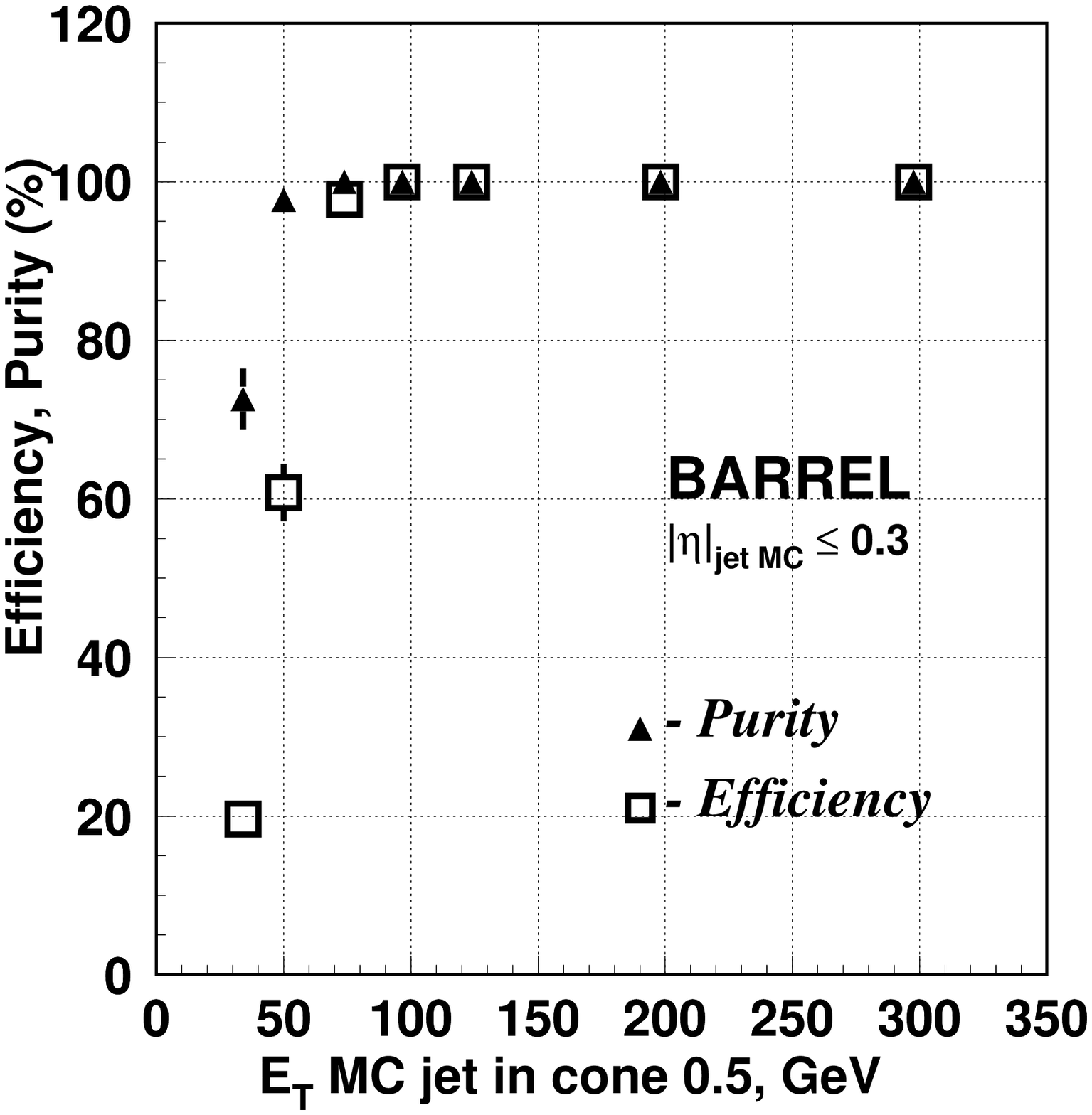}

  \caption{\label{fig:jet} Reconstruction efficiency and purity of jets in
the barrel.}

 \end{minipage}
\end{figure}

\section{Physics studies}

\subsection{Soft}

The tracking of charged particles using the silicon tracker is efficient
and results in low fake rates for \pt$>$ 1 GeV/$c$
(Fig.~\ref{fig:tracking}). An efficiency of 80\% at $dN/dy = 5000$ is
expected even for tracks in jets. The relative resolution of \pt~is about
1\% for tracks in the 1-30 GeV/$c$ range. In order to achieve these
results a rethinking of the reconstruction algorithms and tuning of their
parameters was necessary.

One of the first results of the heavy ion program could be the measurement
of charged particle multiplicity using the silicon pixels alone
(Fig.~\ref{fig:multi}).

Using the electromagnetic and hadronic calorimeters, azimuthal asymmetry
and flow effects can be studied.

\subsection{High \pt}

The extraction of the quarkonia signal is possible via the decay to
$\mu^+\mu^-$:  the CMS detector is well suited for detection of muons up
to 2.5 in pseudorapidity. The mass resolution of dimuons in the $\Upsilon$
region is 50 MeV/$c^2$ (Fig.~\ref{fig:quarkonia}) \cite{bedjidian}, as
compared to about 100 MeV/$c^2$ in ALICE. The background comes from
$\pi/\mathrm{K} \rightarrow \mu$ decay and via $\mathrm{c\overline{c}}$ or
$\mathrm{b\overline{b}}$ production.

One of the strong points of the CMS Heavy Ion program is the detection of
high energy jets, using calorimetry \cite{lokhtin} and also tracking.  
Special jet finding algorithms have been developed which are able to work
in the high background of a central A+A collision (subtraction of
background, contribution from tracks, counting energy in a cone). High
purity and reconstruction efficiency can be reached for jets with \Et~$>$
50 GeV (Fig.~\ref{fig:jet}). Good linearity is seen with a jet energy
resolution of 16\% at 100 GeV.

\begin{figure}
 \begin{minipage}[b]{0.525\linewidth}
  \centering
  \includegraphics[width=\linewidth]{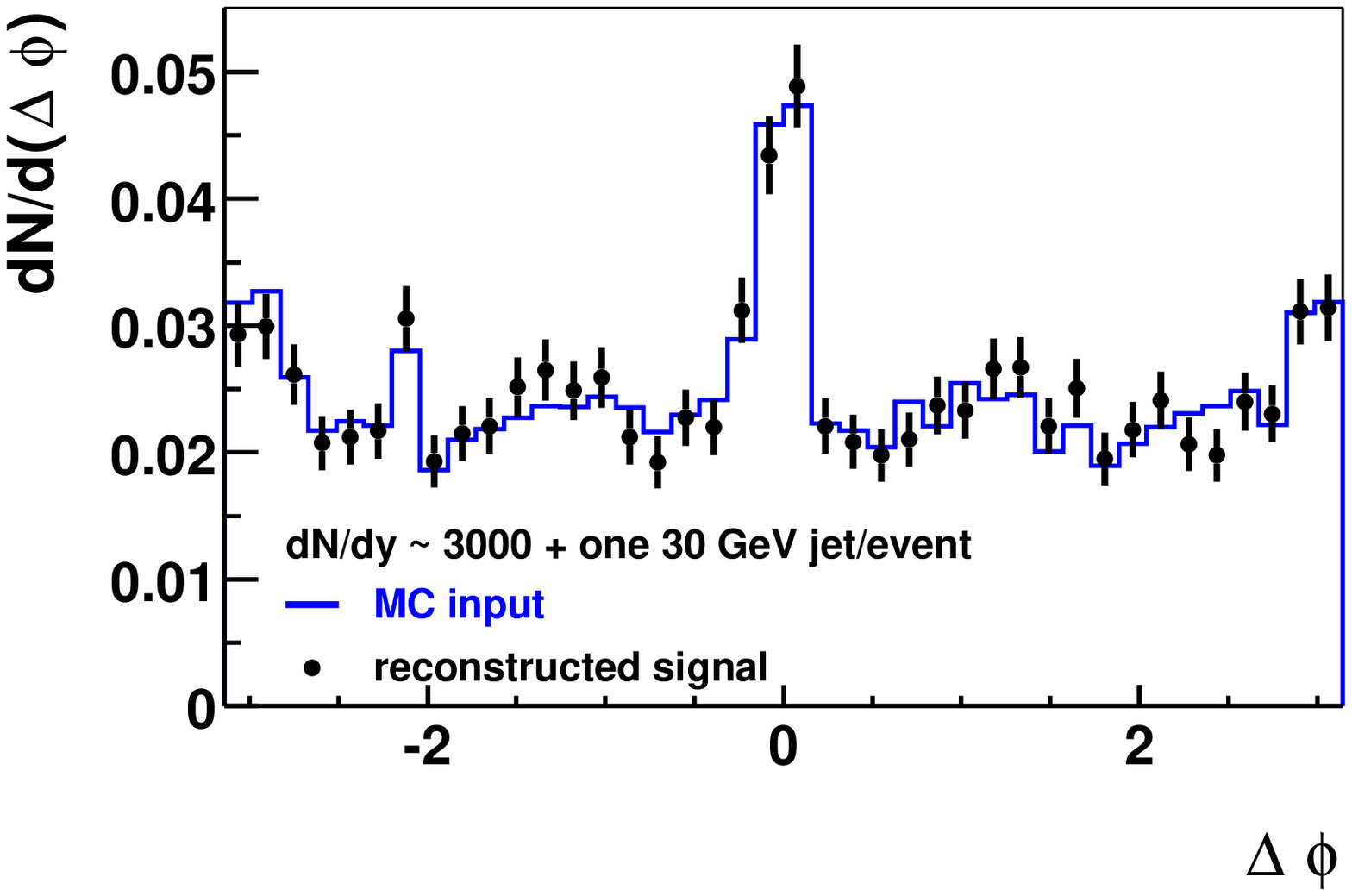}                    

  \caption{\label{fig:azim} Azimuthal correlation of a 30 GeV/event jet on
a $dN/dy \sim 3000$ background: simulated and reconstructed signal.}

 \end{minipage}
 \begin{minipage}[b]{0.475\linewidth}
  \centering
  \includegraphics[width=\linewidth]{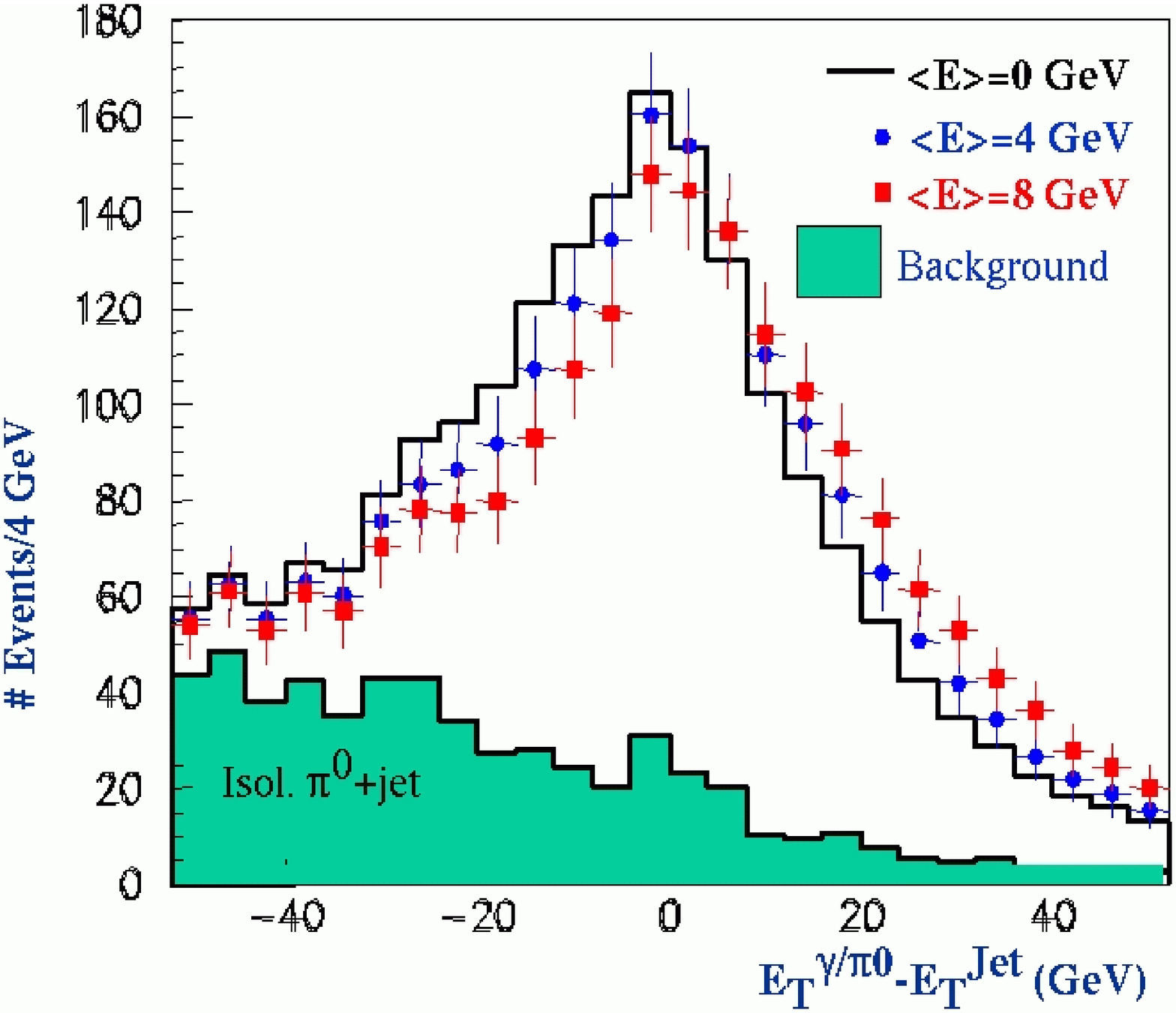}            

  \caption{\label{fig:jetcorr} Distribution of energy difference between
the jet and its photon partner at several average energy loss values.}

 \end{minipage}
\end{figure}

Correlation of jets can be studied as well: good angular resolution is
expected (Fig.~\ref{fig:azim}).

Due to the higher available energy, the correlation of the jet with a
weakly interacting partner can be observed: jet-$\gamma$ and
jet-$\mathrm{Z^0}$ (Fig.~\ref{fig:jetcorr}). This way the energy loss and
partonic propagation effects of the individual jets through dense matter
can be directly measured and compared to predictions: the back-to-back
photon or $\mathrm{Z^0}$ should give the unaffected \pt~of the jet at the
point of its initial production.

\section{Summary}

The CMS Heavy Ion group is a growing international collaboration with an
exciting physics program. Topics to be studied include particle
multiplicities, spectra and correlations, with an extensive high
\pt~program looking at quarkonia, medium effects on jets and jet
correlations.

\section*{Acknowledgment}

The author wishes to thank to the Hungarian Scientific Research Fund
(T 048898).

\vfill\eject
\end{document}